\documentclass[%
 aapm,
 mph,%
 prb,amsmath,amssymb,longbibliography,nofootinbib,
]{revtex4-1}

\usepackage{graphicx}% Include figure files
\usepackage{bm}% bold math
\usepackage{amsmath,amssymb}
\usepackage{braket}
\usepackage{hyperref}
\begin{document}

%\preprint{APS/123-QED}

\title{The process of constructing new knowledge: an undergraduate laboratory exercise facilitated by a vacuum capacitor-resistor circuit}% Force line breaks with \\
%\thanks{A footnote to the article title}%

\author{Frank V. Kowalski}
\affiliation{Physics Department, Colorado School of Mines}%Lines break automatically or can be forced with \\

 %\email{Second.Author@institution.edu}

\date{\today}% It is always \today, today,
             %  but any date may be explicitly specified

\begin{abstract}
The process of constructing knowledge is typically taught to students by having them reproduce established results (e.g., homework problems). An alternative pedagogical strategy is to illustrate this process using an open problem, such as voltage decay in an RC circuit as described below. Analyzing data from this circuit in an undergraduate physics laboratory course reveals a discrepancy between the data and the exponential decay model found in textbooks. As students attempt to reconcile this discrepancy, the instructor can provide guidance in the process of validating data, modeling, and experimental design. This undergraduate laboratory exercise also provides an engaging transition from classroom learning to real world experience. 
\end{abstract}

%Analyzing data from this circuit in an undergraduate laboratory provides an opportunity for the student to engage in hypothesis construction, modeling, and experimental design in attempting to reconcile the discrepancy between the data and the$RC$decay model derived from Kirchhoff's law.

\pacs{Valid PACS appear here}% PACS, the Physics and Astronomy
                             % Classification Scheme.
%\keywords{Suggested keywords}%Use showkeys class option if keyword
                              %display desired
\maketitle

%\tableofcontents

\section{\label{sec:level1} Introduction}

The process of constructing new knowledge in physics may best be demonstrated with an open problem, rather than with a problem whose solution is found in a textbook, on the web, or by using artificial intelligence software. This is particularly difficult in the undergraduate laboratory setting, where student understanding and budget constraints limit the sophistication of the problem and the apparatus. One strategy that can be used is to revise a typical first year laboratory experiment by improving its accuracy, since the validity of every scientific model is in this manner challenged.

Unlike the skills students acquire in a first year course applying Newton's laws to solve a problem, there exists no recipe for creating models that support the data in an open problem. It is not within the scope of this paper to attempt to describe such a generic method to attack an open problem in physics;  rather, a process is outlined for refining understanding of one example, an RC circuit. The purpose here is to provide a simple illustration of scaffolding student knowledge construction in a undergraduate laboratory setting. This is not a new pedagogical approach. It is found in senior projects and graduate school. Although applying it earlier in the curriculum is unusual, this may provide students with beneficial exposure into how knowledge is constructed by physicists.

Different aspects of the results presented below were effectively used to facilitate learning in a junior level advanced laboratory course (over 200 students during a 5 year span) and senior design projects. The emphasis in the former course was on scientific writing, experimental techniques, and critical thinking about data used to support models while that in the latter was on the processes of modeling and experimental design.

%describe the circuit then data then say there are two knowledge issues (1) is the data a result of the experiment not working as expected and (2) is the theory not appropriate. first focus on (1) discuss experimental design associated with the models students construct. finally justify that this is an open problem. finish with curiosity

\section{\label{sec:expt} Methods}

\begin{figure}[ht]
\includegraphics[width=.9 \columnwidth]{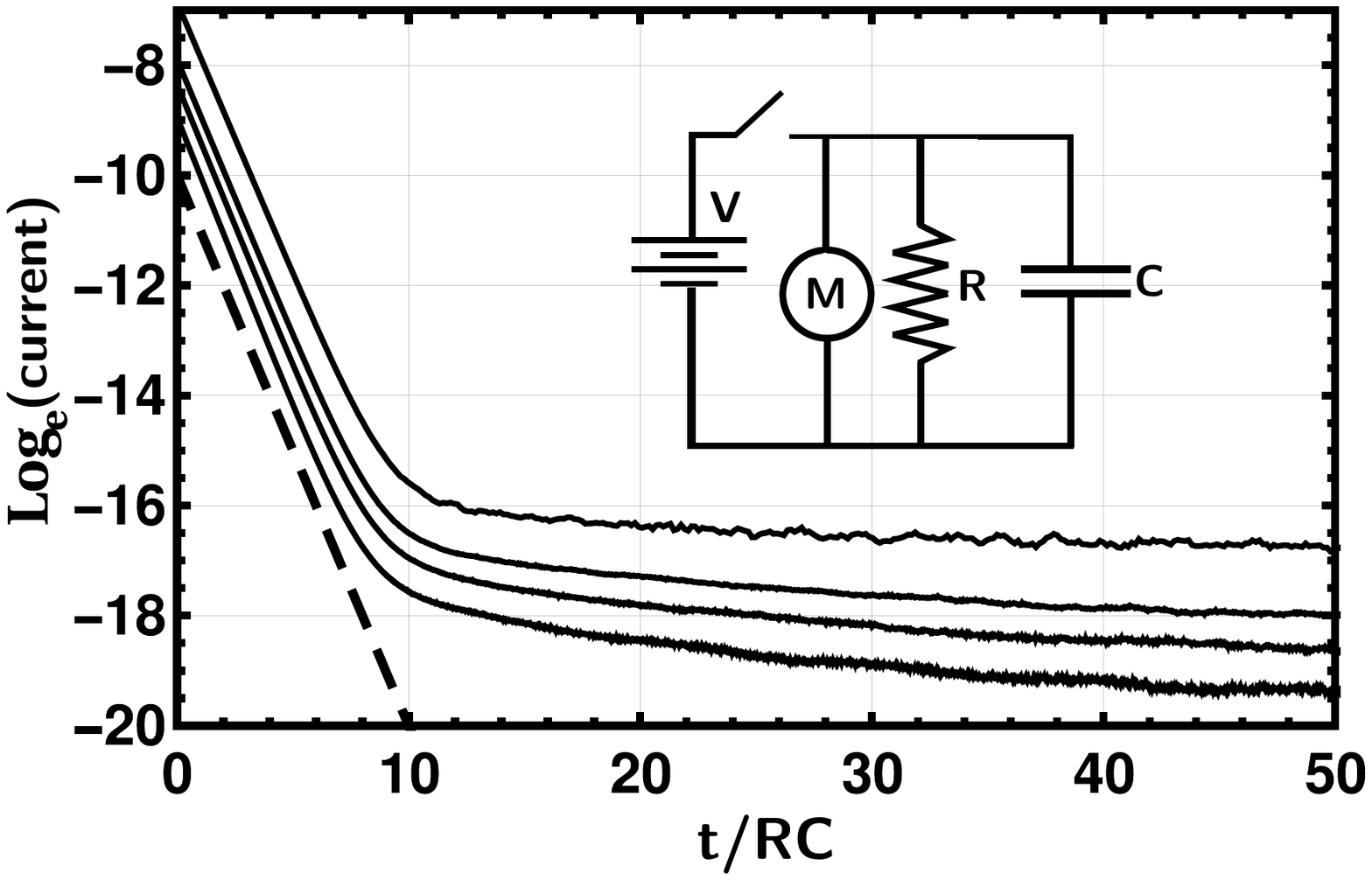}\hfill
\caption{Logarithm of the current vs time in time constant units for the circuit in the inset. A vacuum capacitor is used while the the lines from top to bottom correspond to data for increasing resistances. The expected RC decay for the lowest trace is shown as the dashed line. It is offset for clarity.}
\label{fig1}
\end{figure}

The RC circuit shown in fig.~\ref{fig1} consists of a parallel resistor and a vacuum capacitor with $C=2.789$ nF (Comet model CFMN-2800BAC/8-DE-G). The $1\%$ metal film resistors are $1/4$ W. A $98$ V power supply energizing the circuit is disconnected using a mechanical switch. The needed improvement in accuracy (over a typical 8-bit laboratory oscilloscope) to illustrate deviations from single exponential decay is accomplished with a Keysight 34465A digital voltmeter (used to collect all the data shown here).

The current decay in this circuit is shown as the solid lines from top to bottom in fig.~\ref{fig1} corresponding to circuit resistances $R=47~\mathrm{k}\Omega$, $R=182~\mathrm{k}\Omega$, $R=337~\mathrm{k}\Omega$,  and $R=637~\mathrm{k}\Omega$. In clear contrast to the predictions of Kirchhoff's laws these data deviate from the expected single exponential decay that is shown as the dashed line.

%linear plot shows current reversal and non exponential decay

\section{\label{sec:knowledge} Constructing knowledge}

Creating knowledge about this circuit can be classified into two parts. The first involves determining that the results are repeatable and known with certainty to reflect the behavior of the RC circuit; that is, establishing confidence that the results are valid. The second involves creating a model that the data support within error.

\subsection{\label{sec:knowledge expt} Establishing experimental validity}

%(Hammarlund model 2716-15)

%two oscilloscopes (Picoscope models $4262$ and $4226$) with either $16$ bit resolution at $10$ MS/s or $12$ bit resolution at $50$ MS/s, along with a digital voltmeter (Keysight 34465A). The oscilloscope probes were set on a x$10$ scale with an input resistance $10$ M$\Omega$ and a capacitance of $15$ pF while the digital voltmeter was set to an input impedance $> 1$G$\Omega$.

One method to determine data validity involves questioning the function and influence of each system component. For example, the switch not only disconnects the conducting path but thereafter it has an intrinsic capacitance. The wire loop of the circuit generates a self-inductance and therefore an electromotive force during the decay. The capacitor response is often modeled as having capacitance, series resistance, and inductance, the effects of which need to be considered (self resonance typically occurs at frequencies above $100$ MHz \cite{murata}). The resistance varies with temperature, which is a function of the electrical power dissipated in the resistor (which is a function of time during the decay). The capacitor constricts as a function of voltage, thereby changing its value. The analog-to-digital converter behaves non-linearly at some level of accuracy. Aware of the potential influence of each of these system components, the student can at this point design experiments or perform calculations to determine the impact of these effects (none of which invalidate the data presented).  

However, a major issue in acquiring data that is precise and valid is associated with the small capacitance values that are found in vacuum or air capacitors. Since the resolution of digital voltmeters and oscilloscopes decreases with increasing bandwidth the student may at first attempt to employ large resistance values in the RC circuit to increase the time constant and therefore decrease the bandwidth associated with the decay. It soon becomes apparent that this increases the background noise. One solution for fixed background noise is increasing the energizing voltage. However, the student must then consider safety issues and design the circuit with resistors that have the appropriate wattage.

High bandwidth oscilloscopes with $16$ bit resolution at $10$ MS/s are available (e.g. Picoscope model $4262$). However, even for RC circuits with ten microsecond time constants the frequency response of the oscilloscope probe can introduce oscillations in the decay data or, due to its input impedance, load the circuit. Such issues of data validation can be addressed by collecting data from two probes that are placed at the same point in the circuit. Different data in this case is often the consequences of improperly compensated probes while loading of the circuit by the probe impedance changes the magnitude of the measured responses between using one and both probes.

This RC laboratory exercise exposes students to generic experimental techniques that are often needed to collect reliable and repeatable data. Examples of the latter are the advanced triggering of an oscilloscope that is needed to mitigate the effect of switch bouncing. \cite{bounce} The use of a shielded metal box to mitigate noise is another important method. Taking multiple samples and averaging over these is yet another method. Setting the data acquisition rate much higher than the time constant and using a moving average on the logarithm of the data effectively introduces a low pass filter that also increases precision. Using larger capacitance polypropylene capacitors to collect decay data with much larger time constants illustrates the advantages of power line period averaging available in many digital voltmeters. These methods were applied in collecting the data presented here.

\subsection{\label{sec:thinking} Constructing models}

Aware of the discrepancy yet confident of the validity of their data, the student can now be guided by the instructor to construct new models to explain their observations.

\subsubsection{Using analogical models}

The student can be exposed to the use of analogical models in an attempt to provide insight into this circuit behavior. In this regard and in an effort to help the student make connections, the instructor may first challenge the student to suggest physical processes that exhibit exponential decay and then to consider how such decays may possibly be modified to exhibit the behavior shown in fig.~\ref{fig1}. \cite{peshkin}

\subsubsection{Fundamental principles}

The student should now clearly recognize that the physics curriculum is focused on understanding fundamental principles rather than on manipulating a litany of equations. When a novel phenomenon is encountered it is considered from the perspective of these principles; this emphasis on fundamental principles should stimulate questions associated with validity of these laws.  

The RC decay is most often modeled using Kirchhoff's and Ohm's laws. This emphasis on fundamental principles should arouse curiosity about the validity of these laws. Although Ohm's law is not fundamental it is expected to be appropriate in the analysis of the circuits described here, while Kirchhoff's laws are strictly valid for steady state behavior. 

In recognizing that the steady state regime may not be appropriate, the student might ask if the data reproduces Kirchhoff's result for much larger time constants. However, that does not appear to be the case for the data shown in fig.~\ref{fig1}.

Kirchhoff's laws are superseded by Maxwell's equations. Only a few circuits, involving steady state behavior, have been treated in this manner analytically. \cite{sommerfeld, heald,chabay,klee,muller,moreau}  Maxwell's equations have been solved numerically for an RC circuit. \cite{preyer} However, the data presented here do not support the results of that calculation. 

The intent of this exercise is not for the student to develop a novel solution to RC decay using Maxwell's equations, but rather to use this open ended problem as an example of how a physicist begins to make sense of such data. What often distinguishes an expert from a novice is knowing what to question in light of confounding data. The experimental results described above allow the instructor to awaken and further encourage such awareness.

\subsubsection{Constructing models that eliminate all extraneous variables}

The essence of a problem can often be revealed by a model that eliminates all extraneous variables (as illustrated by the spherical cow metaphor \cite{wiki}). The decay of a vacuum or air capacitor provides a system where this approach can be fruitful. 

Students can be asked to construct the simplest possible system in RC decay. As an example, they may consider replacing the vacuum capacitor with a single spherical conductor (having only self capacitance) attached to ground (or to infinity) via a resistor through which the charge flows in the decay. This model eliminates the effect of the second plate on the decay and appears to contain the bare minimum number of components needed to generate  RC decay. 

Having established this thought experiment the student may focus on how to implement it: using the metal sphere available from a Van de Graaf demonstration apparatus, or soldering the terminals of the vaccum capacitor together. A careful design is first required to determine feasibility, particularly since the capacitance of the measuring apparatus typically overwhelms the small self capacitance of the sphere or the capacitor with its terminals soldered together.

\begin{figure}[!]
\includegraphics[width=.9 \columnwidth]{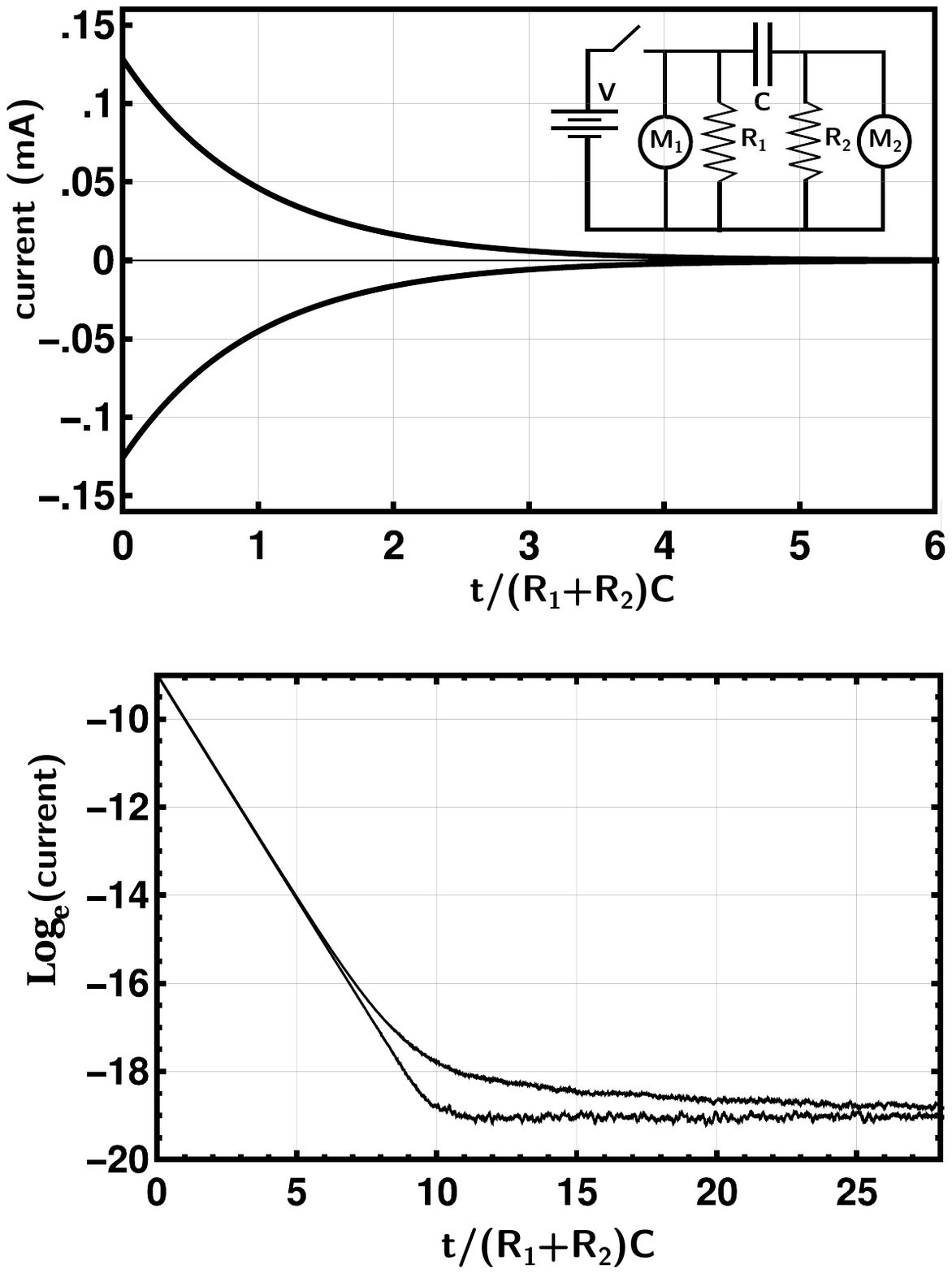}
\hfill
\caption{The decay currents through $R_{1}=337~\mathrm{k}\Omega$ (upper trace) and $R_{2}=337~\mathrm{k}\Omega$ (lower trace) as a function of time. The vacuum capacitor has $C=2.789$ nF. }
\label{fig2}
\end{figure}

To more accurately model the vacuum capacitor the student may choose to add another spherical conductor, connected to ground through a different resistor, and brought near to the first spherical conductor. The system then has both self and mutual capacitance. 

Such a revised thought experiment generates questions about the effect of the charges on one sphere attracting those on the other and thereby modifying the decay process, potentially causing the non-exponential decay in the vacuum capacitor. The charges on one plate repel each other, leading to the voltage decay, while the opposite charges on the nearby plate counteract this decay. 

The addition of this second spherical conductor may lead the student to the idea of connecting each plate of the vacuum capacitor to ground via different resistors. One advantage of including another resistor in the circuit is that the voltage across each resistor reveals the currents flowing into or out of each capacitor plate. The student might then revise the experiment as shown in the circuit schematic of fig.~\ref{fig2}.

\subsubsection{Testing and refining the new model}

A cursory examination of this data in the upper frame appears to support equal currents flowing into and out of the capacitor as expected from Kirchhoff's law. The student may be tempted to go no further in their pursuit of experimental confirmation of the Kirchhoff model. However, when prodded to use a logarithmic scale, discrepancies are apparent, as shown in the lower frame of fig.~\ref{fig2}. This illustrates the importance of attention to detail and how an improvement in precision challenges a scientific model. 

The current from the right capacitor plate almost matches that from a single exponential decay while that from the left plate has a distinct tail. This is evidence for a time varying net charge on the capacitor. However, the initial charge is not measured with this circuit.

The student may choose a different modification of the single charged sphere model: placing a neutral isolated spherical conductor near to the charged spherical conductor. In this case the exponential decay of the latter is modified by the charges induced on the isolated sphere (rather than those drawn to the second sphere through its resistor connected to negative side of the battery). Perhaps the charge on the first sphere does not decay to zero due to the mutual attraction between it and the induced charges on this isolated sphere. 

\begin{figure}[!]
{%
  \includegraphics[width=.7 \columnwidth]{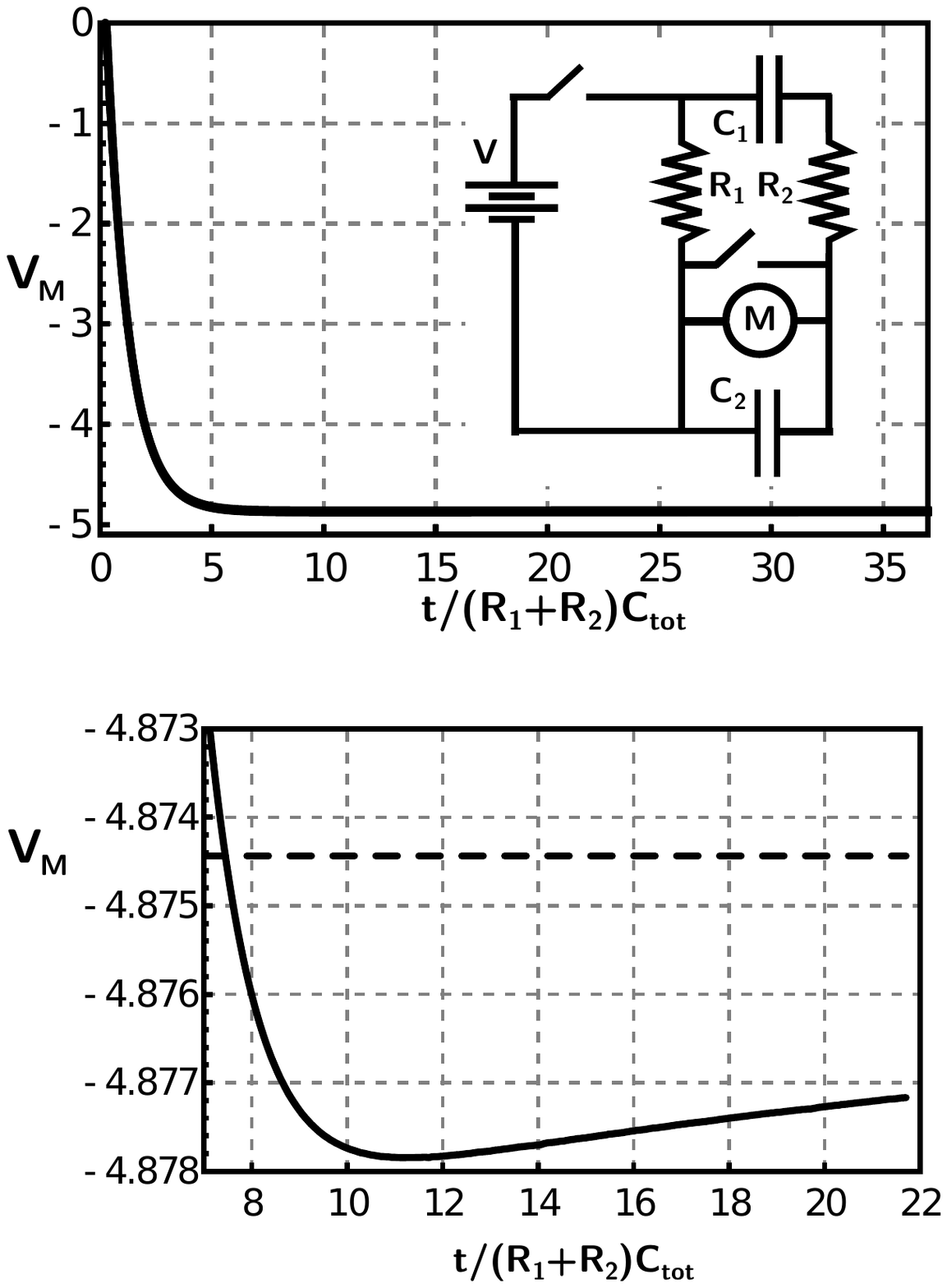}%
}\hfill
\caption{Voltage across $C_{2}$ for resistor values of $R_{1}=1~\mathrm{k}\Omega$ and $R_{2}=337~\mathrm{k}\Omega$ with capacitor values $C_{1}=C_{2}=C=2 \mu$F (film metalized polypropylene axial capacitors) for the circuit shown in the inset of the upper frame. The two switches are initially closed to charge the circuit then open simultaneously. The dashed line indicates the final voltage $V_{f}$. The circuit and data are from reference $14$.}
\label{fig3}
\end{figure}

To test this the student might use two capacitors in series as shown in upper frame of fig.~\ref{fig3}. The right plates are similar to the isolated sphere. By measuring the voltage across $R_{2}$ the current between the right plates is determined (such current data are not shown). 

%In addition, the voltage across $C_{2}$ determines its charge.

\begin{figure}[!]
{%
  \includegraphics[width=.7 \columnwidth]{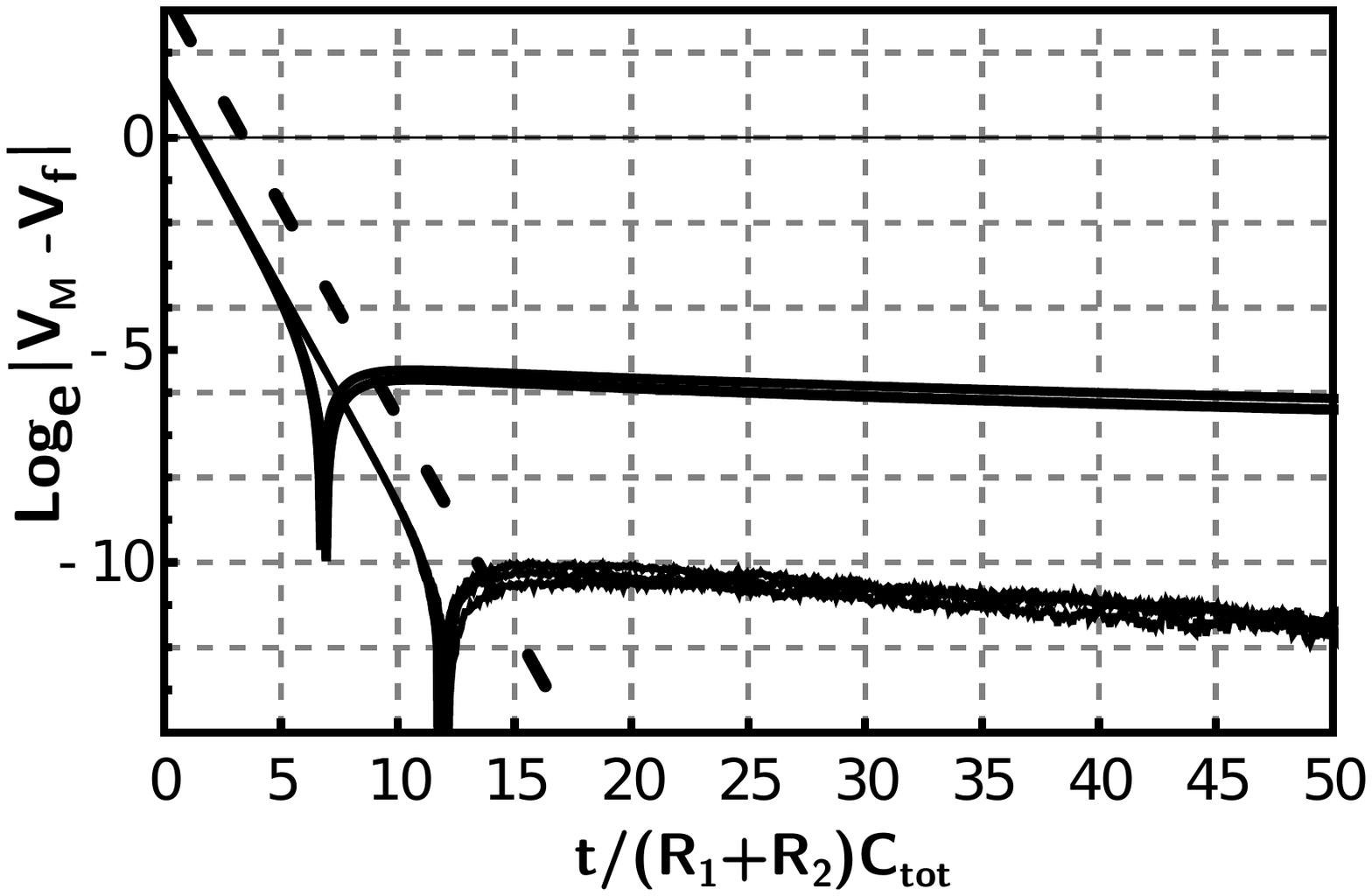}%
}\hfill
\caption{Plots of $\log_{e}|V_{M}-V_{f}|$ for the circuit in the upper frame of fig.~\ref{fig3}. Four decays are are shown when the switch across $C_{2}$ is operational (the four upper traces that appear as one solid line) and when it is always open (the lower four traces).}
\label{fig4}
\end{figure}

Initial net negative charge on the right plate of $C_{1}$ and no net charge on the right plate of $C_{2}$, in the upper frame of fig.~\ref{fig3}, is generated when the two switches are closed. The decay is measured when these two switches are opened simultaneously (within a millisecond using a double pole single throw mechanical switch for a $338$ millisecond time constant). After the two switches open some of the charge on the right plate of $C_{1}$ relaxes to right plate of $C_{2}$. The voltage between the plates of $C_{2}$ decreases below its equilibrium value $V_{f}$, as shown in \ref{fig3} (b) by the dashed line, before returning to it. On a logarithmic scale of the absolute value of $V_{M}-V_{f}$ vs time this minimum results in a dip as shown in fig. \ref{fig4}.

The more familiar initial condition for the series capacitor configuration results from removing the switch across $C_{2}$. For comparison the decay for this initial condition is shown in fig.~\ref{fig4}.

%Therefore, $dV_{M}/dt$ switches sign through this minimum. Decays similar to those shown in fig.~\ref{fig1} occur simply by varying $V$. This should convince the student that inductance does not significantly influence this process.

%The circuits in figs. \ref{fig1} and  \ref{fig3} illustrate both sequential decays with different time constants and with a change in current in the transition regions.

The above data provide direction for constructing models. For fig.~\ref{fig1} the most obvious is $I_{M}=I_{10} \exp[-\alpha t]+I_{20} \exp[-\beta t]$ where $\alpha=1/RC$ and $\beta$ is the second decay constant. These parameters for the data of fig. \ref{fig1} are shown in fig. \ref{fig5}. For fig.~\ref{fig3} this becomes $V_{M}=V^{in}_{f} (1-\exp[-\alpha t]) + V^{out}_{f} (1- \exp[-\beta t])$ where $V_{f}= V^{in}_{f}+ V^{out}_{f}$. This is referred to as the sum of exponentials model (fits typically overlap the data presented here within the width of the lines in the figures).\cite{kowalski}

Plots of current vs time in time constant units allow for a range of data to be presented on one graph. However, the model parameters must be determined from voltage vs time data.

\subsubsection{Validity of model parameters: dynamic capacitance}

One aspect of modeling is determining the validity of the parameters used. Consider capacitance, for example. The assertion that the ratio of the charge to the potential difference is a constant can be tested with the circuit of fig.~\ref{fig1} by integrating the current leaving one capacitor plate to determine its remaining charge at a given time. Dividing this by the voltage across the capacitor at that time does not yield a constant value throughout the decay. Therefore, capacitance may not be a useful circuit parameter. 

%Reference: https://www.physicsforums.com/threads/proof-of-q-cv-for-arbitrarily-shaped-capacitors.1014957/

However, in modeling non-linear systems a connection is often made between a static parameter and its dynamic version. Examples are found in both static and dynamic resistance (applied to a p-n junction, that was perhaps discussed in an earlier course on electronics) and dynamic inductance (ferromagnetism is often discussed in a first year course). Even the relationship between phase velocity $\omega/k$ and group velocity $d \omega/dk$ (discussed in an earlier course on modern physics) is a similar quantity and its derivative. 

Suggesting to the student that the dynamic capacitance $C_{dyn}=dQ/dV$ may be a useful parameter leads to the ordinary differential equation $C_{dyn} dV/dt=-V/R$. where $V$ is the voltage across the capacitor and $dQ$ the change in charge on the left plate of the capacitor in fig.~\ref{fig1}. Kirchhoff's law for this RC circuit is obtained from this relationship when $C_{dyn}\rightarrow C$ (for exponential decay $C_{dyn}= C$). Using this method a first order ODE has a solution that is a sum of exponentials. For example, $C_{dyn}$ can be determined using the expression for the voltage from the sum of exponentials model described above. The value of $C_{dyn}$ is then the capacitance measured by a capacitance meter for $t\ll RC$ that transforms into the much larger value associated with the $\beta$ relaxation time for $t\gg RC$.

\subsubsection{Return to fundamental principles}

It is surprising that the capacitor, whose dimensions do not change appreciably, can have a capacitance that varies as suggested by the dynamic capacitance model. This conundrum may stimulate the student to focus on acquiring a fundamental understanding of circuit behavior using Maxwell's equations before addressing the data. Such solutions typically involve steady state behavior, utilizing a surface charge on a wire to confine its current. \cite{muller,sommerfeld, heald, chabay,moreau} A numerical calculation involving relaxation in an RC circuit yields an initial non-exponential decay due to transit time effects but does not match the above data. \cite{preyer}

Surface charge effects are demonstrated with circuit wire that has a kink. If the current into the kink is greater than that out,  ``Then the charge piles up at the `knee,' and this produces a field aiming away at the kink. The field opposes the current flowing in (slows it down) and promotes the current flowing out (speeding it up) until these currents are equal, at which point there is no further accumulation of charge and equilibrium is established.'' \cite{griffiths2} 

Drawing an analogy between the charge on the kink in a wire and the capacitor in the circuit of fig.~\ref{fig1} the student may divide the charge into two parts: one that generates a field only between the capacitor plates, $Q^{in}$, that exerts no force on the current in the wires and one that confines the current to flow in the wires, $Q^{out}$. This is referred to as the sum of charges model. 

For example, before the switches are opened in the circuit of fig.~\ref{fig3}, $Q^{out}$ generates zero electric field in the wire segment from the upper node along the branch to the capacitor while the field is non-zero in the segment from the upper node along the other branch to the resistor, thereby directing the current from the battery into $R$ and not to the capacitor. The response due to $Q^{in}$  and $Q^{out}$  is the sum of their individual responses from the superposition principle and is manifest as the first (given by the standard solution for the RC circuit) and second terms in the sum of exponentials model.

\begin{figure}[!]
{%
  \includegraphics[width=.7 \columnwidth]{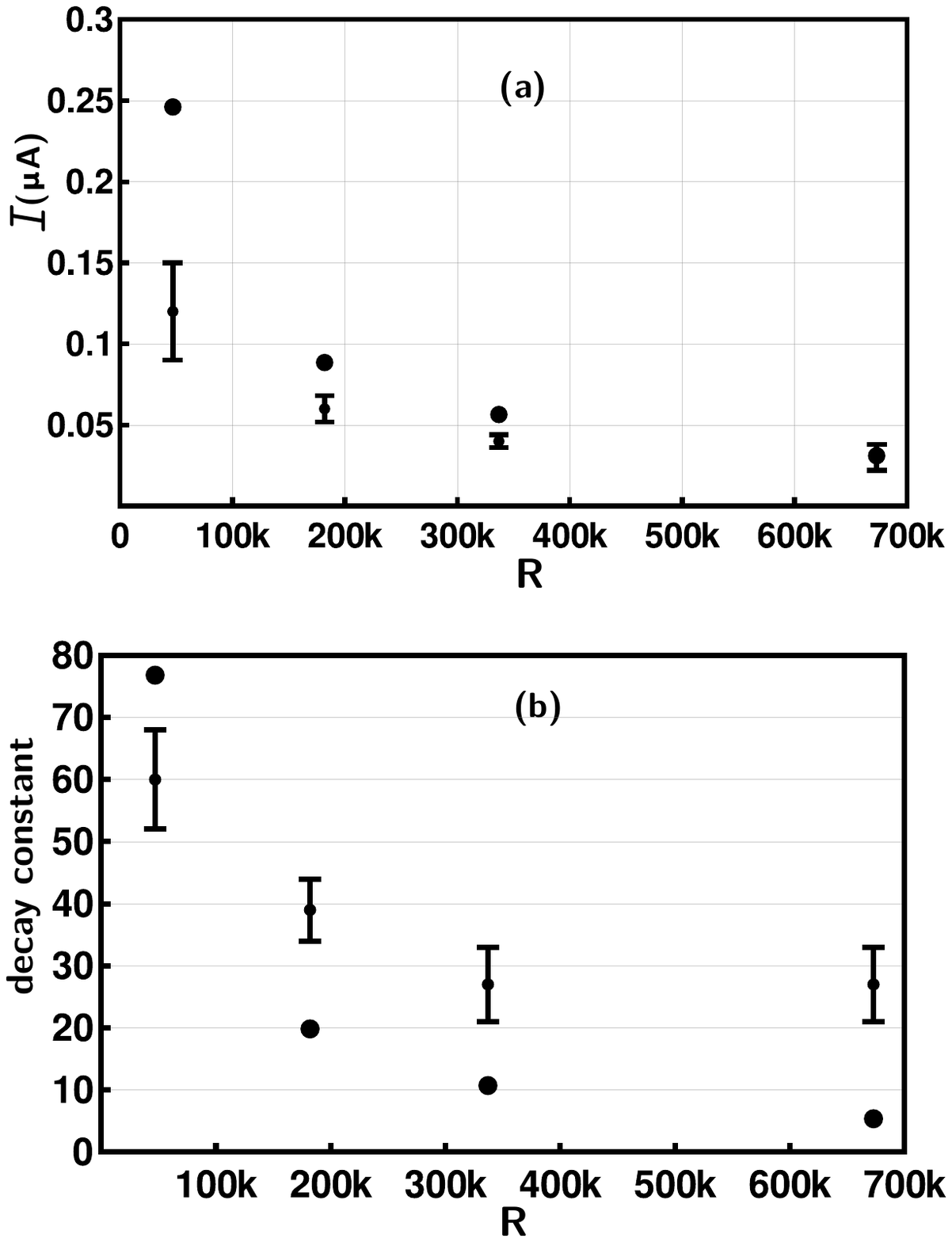}%
}\hfill
\caption{Parameters of the sum of exponentials model as a function of R determined from a fit of the data in fig. \ref{fig1}. The data with error bars in frame (a) corresponds to $I_{20}$ while the solid circles indicate $I_{10}/4200$ (note that this data is proportional to $1/$R).  The points with error bars in frame (b) are the $\beta$ values while the solid circles are the corresponding values of $\alpha/100$.}
\label{fig5}
\end{figure}

The student may attempt to extend the kink in a wire analogy to the RC circuit in fig. \ref{fig1}: a larger charge may be needed to redirect a larger current from the capacitor plate into the resistor. The student might then conjecture that this charge results in an inverse relationship between $I_{20}$ and the resistance in the circuit. The student can check this in a plot of $I_{20}$ as a function of R and compare it to $I_{10}\propto 1/R$ as illustrated in fig. \ref{fig5}.

%Let the parameters of the second term be $Q^{out}_{0}=V^{out}_{0} \tau_{out}/ R$, where $\tau_{out}$ is a constant and $V^{out}_{0}$ is the voltage at the plate generated by $Q^{out}_{0}$. This inverse dependence on $R$ can be related to the kink in the wire; larger charge is needed to redirect a larger current from the capacitor plate into the resistor. This implies that the initial current also has an inverse relation for with R. 

%Support for this conjecture along with other examples of the sum of exponentials and sum of charges models (along with fits of the data) are found in the literature. \cite{kowalski}

Another method to test the sum of charges model is to vary the amount of $Q^{out}$ compared with $Q^{in}$ on a capacitor plate. To do this the student might modify the circuit in fig.~\ref{fig2} by adding a different voltage source, $V_{2}$, across $R_{2}$ using a switch that opens synchronously with the other switch. As $V_{2}$ approaches $V$ the field inside the capacitor is reduced (as is $Q^{in}$) while $Q^{out}$ on the plates has not been diminished (it is still required to direct current through the resistors).  Data for such a modification are also found in the literature.\cite{kowalski}

The sum of charges model attributes the unusual behavior of $C_{dyn}$ to the $Q^{out}$ charges on the capacitor (and the circuit wires) that are not included in the definition of capacitance, $C=Q^{in}/\Delta V$. Since $Q^{in}$ and $Q^{out}$ decay at different rates the capacity to hold charge at a given voltage is not constant. 

%One disadvantage of dynamic capacitance is that a different $C_{dyn}$ is needed for circuits with different values of $R$, $C$, and $V$.

%The series capacitor circuit demonstrates that initial net charge on the right plates of $C_{1}$ and $C_{2}$ accentuates the overshoot of charge compared with no net initial charge on these plates.

\subsubsection{Modeling with a dimensionless variable}

Decay data expressed in terms of the dimensionless variable $\tau=$ t/RC facilitates the determination of the variables that influence the decay. For example, if the relaxation is a function of only R and C then a point on the decay plot at a particular value of $\tau$ will be the same for different combinations of R and C that yield this value of $\tau$ (apart from variations due to the initial voltage). The decay plot is therefore the same for various R and C values. That is the case for the faster but not slower decay shown in fig. \ref{fig1}. Therefore a model that this data supports must include more variables than R and C. From the sum of charges model the student might conjecture that $Q^{out}$ and its distribution on the outside of the circuit wires and capacitor plates is such a variable. 

\begin{figure}[!]
{%
  \includegraphics[width=.7 \columnwidth]{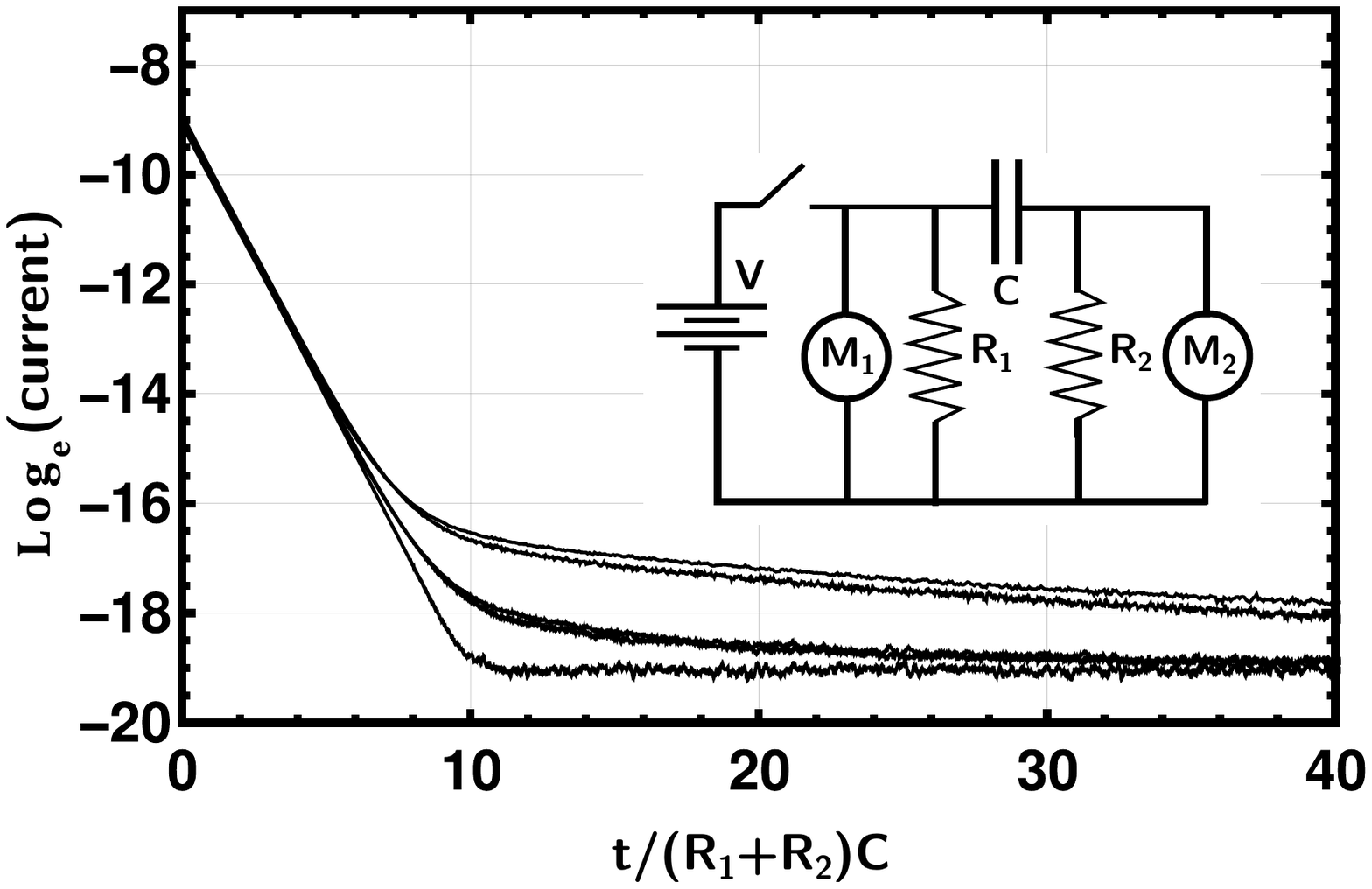}%
}\hfill
\caption{Current relaxation for the circuit in the inset for fixed resistance $R_{1}=R_{2}=337.8~{\textrm k}\Omega$ and different values of the capacitance. The trace just above the bottom trace consists of two decays that overlap. The bottom trace corresponds to all currents through $R_{2}$. A $98$ V power supply was used to energize the circuit. The circuit and data are from reference $14$.}
\label{fig6}
\end{figure}

Fig. \ref{fig6} illustrates the decay as a function of t/RC while C varies, where R$=R_{1}+R_{2}$, for fixed $R_{1}=R_{2}=337.8~{\textrm k}\Omega$. The data are obtained using four vacuum capacitors, two with C$=0.5177$ nF and two with C$=2.789$ nF. The upper trace is the current through $R_{1}$ with C$=0.5177$ nF and the trace just below it (that almost overlaps with the upper trace) is the current through $R_{1}$ for C$=1.0354$ nF (a parallel combination of two C$=0.5177$ nF capacitors). The next two traces below these are for the current through $R_{1}$ with C$=2.789$ nF and C$=5.578$ nF (a parallel combination of two C$=2.789$ nF capacitors). These traces completely overlap. The lowest trace is the decay data for the current through $R_{2}$ for all four values of C.

Consider the above discussion about the determination of variables in plots of the decay as a function of $\tau$ for the upper pair of the traces that are assumed (for the sake of argument) to overlap. The doubling of C does not change the plot (it of course does in a plot vs t). This implies that R and C are the only variables required to model the data (assuming for the moment that variation in R also yields the same plot).

Similarly, the overlap of the plots of the current decay vs $\tau$ for C$=2.789$ nF and C$=5.578$ nF implies that R and C are the only variables required to model the decay from these capacitors (again assuming for the moment that variation in R also yields the same plot). However, a new variable must exist to model both pairs of data since the decay differs for the two pairs of lines. One then looks for a variable that is the same for each pair of capacitors but differs between them. Their geometry is such a variable. From the sum of charges model the student might again propose that the distribution of $Q^{out}$ due to this geometrical difference is the needed variable and then propose experiments to test this hypothesis.

%The graphs for one cap and two of the same capacitance in parallel vs time in time constant units should be the same for all R and C values since R and C are no longer parameters in such plot. Indeed, they are the same for one cap pair and the same for another cap pair but differ when using a different cap pairs with different capacitances but with the same resister. In such a graph the variables R and C are eliminated. That is you don't have to plot decay vs R and then decay vs C since R and C are no longer variables in the equation. This reduces a plot of decay in two dimensions to only one dimension. If there is a difference in these plots then it to do with a different variable. In this case of pairs of caps that generate different plots it is the geometry of the capacitors. If the graphs vary with R on a plot in time constant units then a different variable is at play.
%\https://physics.stackexchange.com/questions/131732/why-is-it-natural-to-look-for-solutions-involving-dimensionless-quantities

\section{Summarizing the new knowledge constructed}

The knowledge gained from the above exercise is divided into two parts: the knowledge acquired in obtaining precise and valid data (experimental techniques that are typically unknown to the student but not novel) and the consequences of the models that these data support. 

\subsection{Knowledge the student gains in constructing precise and valid data}

\begin{itemize}
\setlength\itemsep{-0.5em}

\item Advanced triggering methods (often available in an oscilloscope) required to obtain repeatable data; \item The technique of averaging these repeatable data sets to improve precision; \item Filtering techniques in the form of a moving average over the oversampled data to increase precision; \item Shielding methods to mitigate noise (including a properly grounded metal enclosure of the circuit); \item The technique of increasing the signal but not the noise (by increasing the energizing voltage); \item Safety and design considerations in applying a high voltage energizing source to the circuit;\item Understanding that high impedances increase noise pickup; \item Methods of validating the data such as placing two probes at the same point in the circuit to determine the effect of their impedance on the circuit or to determine if the probes have different frequency responses; \item Use of power line period averaging available in digital voltmeters to mitigate power line noise.

%\item The need for a voltmeter with greater than $10^{9}~\Omega$ input impedance to prevent loading the circuit in fig. \ref{fig3}.

\end{itemize}

\subsection{Knowledge gained in modeling this data}

\begin{itemize}
\setlength\itemsep{-0.5em}

\item At least two relaxation mechanisms occur during the decay with disparate time constants. The first decay constant is that expected from Kirchhoff's laws while the second is unrelated to the first; \item Unequal currents flow into and out of the capacitor in RC decay; \item Net charge is generated on the capacitor; \item The sum of exponentials model is supported by the data; \item The series capacitor circuit provides a method to explore how charge flows between isolated conductors during the decay; \item Valid data for RC decay, obtained for dielectric capacitors in series, indicate a change in current direction during the decay; \item The use of dynamic capacitance in modeling this data is of limited utility since it does not present a microscopic understanding of the process and therefore has little predictive power; \item The sum of charges model supports the sum of exponentials model since the initial decay time constant, RC, is the same for all data sets while the parameters of the the second term in the sum of exponentials model need to be adjusted to match the particular decay. For example, $Q^{out}$ for the left plate in fig.~\ref{fig2} differs from that on the right plate in order for the sum of exponentials model to match the data from each plate, yet the initial decay term for both plates involves only the time constant $(R_{1}+R_{2})C$. \item Before the switch is opened $Q^{out}$ directs current into $R$ while the electric field from $Q^{in}$ has no such influence since its field is confined to a region between the plates (without $Q^{out}$ and before the switch is opened the student might ask what force causes current to flow through the resistor if the field from $Q^{in}$ does not exist in the wires to move charges); \item The description of transient behavior using a kink in a wire is related to the behavior of the capacitor and consistent with the sum of charges model;\item Exponential decay of charge is the result of $dQ/dt \propto -Q$. However, the decay constant that makes this relationship an equality is determined by the details of the decay mechanism which requires a solution of Maxwell's equations for the circuit; \item The decay of the circuit in fig.~\ref{fig2} with a different voltage source across each resistor (that initially energizes it) allows for a variation in $Q^{in}$ while $Q^{out}$ is essentially fixed, thereby facilitating tests of the sum of charges model; \item Plots in terms of dimensionless variables demonstrate that the geometry of the capacitor is an important variable in the decay process; \item While progress has been made a complete microscopic understanding of the decay process, based on Maxwell's equations, is lacking.

\end{itemize}

As with any model questions arise that probe its consequences. The student may wonder about the effects of $Q^{out}$ in an inductor-resistor circuit, or about the consequences of the sum of charges model when the loss mechanism involves transfer of the electrical energy stored in the capacitor to mechanical energy or to other circuit components. Perhaps a superconducting LC circuit, whose loss mechanism is radiative rather than thermal, behaves in the same manner. On a theoretical level, they may wonder why the simple sum of exponentials model matches the data even though it is quite difficult to obtain such a solution from Maxwell's equation. 

Through this exercise, the student not only has experienced the process of constructing new knowledge, but also experienced the questions and uncertainties it often exposes.

%Does a charged spherical conductor decay with a single exponential or do the charges on the wires also generate a sum of exponentials decay?

\section{\label{sec:conclusion} Conclusions}

%the physics seems simple to understand: the attraction between plus and minus charges. However, there are subtle issues for example the crawford comment that the uniqueness theorem forces only one solution that goes against intuition for mutual attraction. the mathematics of applying maxwells eqns is difficult.

Students in an upper-level undergraduate physics laboratory class can be effectively exposed to the process of constructing new knowledge in physics. This involves the acquisition of valid data (whose importance is emphasized in this procedure) along with the construction of models supported by the data within error. The simple RC circuit, for which accurate measurements can be made inexpensively, is surprisingly difficult to model on a fundamental level. Yet the physics involves the familiar attraction and repulsion between charges. The students should be cognizant of the learning objective: the point is to learn how to collect valid data, practice modeling, and think critically about the data supporting the models, not to find a unique model that the data support. Hopefully, joy can be found in this process rather than only in the result, for which complete success is often elusive.

This laboratory experience also provides an opportunity for students to become aware of the differences between the simplified models ubiquitous in their formal education and the more complex realities of the physical world. Not only do students gain fresh insight into how new knowledge is constructed, but they move toward a more mature and sophisticated view of physics.

\begin{acknowledgments}
I wish to thank Susan Kowalski for editing comments from a non-physicist's perspective, as well as students for their participation and discussions in a senior design course, particularly Justin L. Swantek, Tony D'Esposito, and Jacob Brannum. The support from a HP Technology for Teaching Grant is acknowledged.
\end{acknowledgments}

\section{\label{sec:appendix} Appendix}

The instruction described above is a fruitful but time intensive endeavor. Alternately, a shorter exercise described next saves considerable course time while still cultivating some skills essential for constructing new knowledge. This has been successfully implemented as a pre-post assessment in a junior level modern physics laboratory course.

Students are told that the purpose of the pre-test is to determine their skills in performing experiments and communicating their results. Rather than being given written instruction, they are simply told to use the equipment at hand to collect RC decay data and write a short scientific report on their work. 

The data that they collect using a microfarad ceramic or electrolytic capacitor and a $1~\mathrm{k}\Omega<R<100~\mathrm{k}\Omega$ resistor with an 8-bit oscilloscope appears to match Kirchhoff's law during the initial decay. However, its tail deviates significantly from exponential decay even when analyzed with an 8-bit resolution oscilloscope. In the pre-test many of the students conclude that this data indicates exponential decay, verifying what was taught in a previous class (many plot the data on a linear scale). A remarkable number of reports do not reflect an awareness that modeling is a fundamental part of a scientific method; that is, the reports lack an argument that the data support a model within error. 

As a summative assessment later in the semester, the student repeats the experiment (or is given a file of decay data with unique RC values for each student to mitigate cheating).  From this data the student writes a laboratory report. Learning in the course is illustrated as the student compares their pre-test and post-test documents. Although this alternate exercise saves considerable course time, it has the pedagogical shortcoming of not allowing the students practice in modeling and experimental design. Nevertheless, it has the advantage of challenging the student to think critically about the validity of the data, particularly in the case where the student expects the data to support the model within error. This is an important first step in constructing new knowledge.

\end{document}